\documentstyle[epsfig,aps,12pt]{revtex}
\begin{document}

\title{ $\eta'-g^*-g$ Transition Form Factor with Gluon Content 
  Contribution Tested }
\author{ Taizo Muta and Mao-Zhi Yang\\
       Department of Physics, Hiroshima University, Higashi-Hiroshima,\\
       Hiroshima 739-8526, Japan{\thanks{mailing address}}}
\maketitle
{\flushleft PACS Numbers: 12.38.Bx}
\begin{center}
\begin{minipage}{120mm}
\vskip 0.8in
\begin{center}{\bf Abstract} \end{center}
{We study the $\eta'-g^*-g$ transition form factor by using the $\eta'$ wave 
function constrained by the experimental data on the $\eta'-\gamma^*-\gamma$ 
transition form factor provided by CLEO and L3 . We also take into account the 
contribution of the possible gluonic content of the $\eta'$ meson. 
}
\end{minipage}
\end{center}
\vskip 1in

\newpage

The large branching ratios of $B\to\eta' X_s$ and $B\to\eta'K$ detected by CLEO
\cite{1,2} enhanced the importance of the QCD transition form factor 
$F_{\eta'g^*g}\equiv H(q^2_1,q^2_1,m^2_{\eta'})$. The mechanisms using 
this transition form factor to explain these large decay rates are based 
on the $b\to sg^*$ penguin transition followed by the decay
$g^*\to\eta' g$ \cite{3,4,5,6,7,8} or on the transitions
$g^*g^*$, $g^*g\to \eta'$ 
\cite{9,10}. The transition form factor $F_{\eta'g^*g}$ being used are either
extracted from the experimental data of $\psi\to\eta' \gamma$ \cite{3,9,10}, 
or based on 
phenomenological consideration \cite{4,8}, or calculated by assuming 
pseudoscalar coupling between $\eta'$ and a quark pair \cite{5}. Thus 
a question arises, what is the relation between the form factor 
$F_{\eta'g^*g}$ used in the above references and the wave function 
of $\eta'$? Provided that the form factors $F_{\eta'g^*g}$ used in these 
references are not the same, which one can be 
got from the calculation based on the structure of $\eta'$? Thus studying the 
QCD transition form factor $F_{\eta'g^*g}$ is not only important in 
investigating the dynamics of $\eta'$ production from B decays, but also in 
detecting the structure of $\eta'$.

   The $\eta'$ meson is particularly different from the flavor 
octet meson: $\pi$, $K$. It is mainly a singlet meson. According 
to the QCD anomaly it is much heavier than the 
massless Glodstone boson \cite{11}. Because of its singlet structure the 
$\eta'$
meson may have gluonic content. Since two decades ago its gluonic 
structure has been studied in QCD sum rules \cite{12}. Recently the 
experimental data on the 
$\eta' \gamma^*\gamma$ transition form factor from CLEO \cite{13} 
and L3 \cite{14} pushed forward the development of the investigations of the 
quark structure of $\eta'$ \cite{15}. A new $q\bar{q}-s\bar{s}$
(here $q\bar{q}$ means $u\bar{u}$
and $d\bar{d}$) mixing scheme was developed \cite{16}. The calculation based on
the $\eta'$ wave function successfully describes the experimental data on the
$\eta'\gamma^*\gamma$ transition form factor over a wide range of the virtual 
photon momentum squared, $1 GeV^2\leq Q^2 \leq 15 GeV^2$. Here the momentum 
squared of the virtual gluon in the production of $\eta'$ from B decay can 
vary from $1 GeV^2$ to $25 GeV^2$. Is the situation in $g^*g\to \eta'$ or
$g^*\to \eta' g$ transition similar to $\gamma^*\gamma\to\eta'$ transition? 
Maybe not because of the particular QCD structure of $\eta'$. The 
QCD anomaly and gluonic content in the $\eta'$ may play important 
role in the $\eta'-g^*-g$ transition.
The QCD axial anomaly determines the behavior of the $\eta'-g^*-g$ transition
form factor at small momentum transfer, i.e., in the soft limit $Q^2\to 0$.
In the range $Q^2\geq 1 GeV^2$ gluonic content of $\eta'$ may have some 
contributions, which will make difference between the
$\eta'-g^*-g$ transition and $\eta'-\gamma^*-\gamma$ transition.

   In this work we use the wave function of $\eta'$ to calculate the 
$\eta'-g^*-g$ 
transition form factor $F_{\eta' g^*g}$ at the large momentum transfer region
$Q^2\ge 1GeV^2$. We not only take into account the quark content of $\eta'$
but also test how much the gluonic content contributes.

  In the $\frac{1}{\sqrt{2}}\mid u\bar{u}+d\bar{d} \rangle$ and 
$\mid s\bar{s}\rangle$ mixing scheme \cite{16} the parton Fock state 
decomposition can be expressed as
\begin{equation}
 \mid \eta'\rangle = sin\phi \mid \eta'_q\rangle +
                       cos\phi \mid \eta'_s\rangle
                       +\mid G\rangle.   
\end{equation}
where $\phi$ is the mixing angle, and $ \mid \eta'_q\rangle \sim
\frac{1}{\sqrt{2}}\mid u\bar{u}+d\bar{d} \rangle$, 
$\mid \eta'_s\rangle \sim \mid s\bar{s}\rangle$, $\mid G\rangle \sim
\mid gg \rangle$ . In eq.(1)  $ \mid \eta'_q\rangle$ and 
$ \mid \eta'_s\rangle$ are quark Fork states, and $\mid G\rangle$ is
the gluonic Fock state. $\mid G\rangle$ and $ \mid \eta'_q\rangle$,
$ \mid \eta'_s\rangle$ are not independent. Because 
$ \mid \eta'_q\rangle$ and $ \mid \eta'_s\rangle$ are 
non-flavor-octet, the evolution of their wave function will mix
with the two-gluon state. In \cite{17} the evolution equation for the 
wave functions of the mixing $q\bar{q}$ and $g\bar{g}$ state has
been derived. The eigenfunctions have been calculated. After a few
simple steps of algebraic procedures, one finds,
\begin{eqnarray}
   \Psi^q_i(\mu^2,x)&=&f_i\phi^q(\mu^2,x),\nonumber\\
     \phi^q(\mu^2,x)&=&6x(1-x) \left\{ 1+
               \sum\limits_{n=2,4,\cdots}
   [B^q_n\left(\frac{\alpha_s(\mu^2_0)}{\alpha_s(\mu^2)}\right)
    ^{\gamma_+^n}+    
  \rho^g_nB^g_n\left(\frac{\alpha_s(\mu^2_0)}{\alpha_s(\mu^2)}\right)
    ^{\gamma_-^n}]C^{3/2}_n(2x-1)\right\}  \nonumber\\
 \Psi^g_i(\mu^2,x)&=&f_i\phi^g(\mu^2,x),\nonumber\\
          \phi^g(\mu^2,x)&=&x(1-x) \left\{
               \sum\limits_{n=2,4,\cdots} 
 [\rho^q_nB^q_n\left(\frac{\alpha_s(\mu^2_0)}{\alpha_s(\mu^2)}\right)
   ^{\gamma_+^n}+
   B^g_n\left(\frac{\alpha_s(\mu^2_0)}{\alpha_s(\mu^2)}\right)
        ^{\gamma_-^n}] C^{5/2}_n(2x-1)\right\}, 
\end{eqnarray}
where the superscripts $q$ and $g$ indicate the ``quark" and ``gluon" 
content, $i$ denotes the meson state composed of the $q\bar{q}$
pair, and $f_i$ is the decay constant of the relevent meson state
of $q\bar{q}$. The parameter $x$ is the momentum fraction carried 
by the parton. Here $\mu$ is the scale of the hard process, which 
may be taken to be the 
momentum transfer $Q^2$ involved in the hard process. $\gamma_+^n$,
$\gamma_-^n$, $\rho_n^g$, $\rho_n^q$ are not free parameters; they
are given in Appendix A. $C_n^{3/2}$ and $C_n^{5/2}$ are Gegenbauer
polynomials. $\mu_0$ is the reference scale and we take 
$\mu_0=0.5 GeV$ in our calculation. Because of the general symmetry 
properties of the wave functions of the two-particle bound state of a
neutral pseudoscalar meson, the quark wave function satisfies
$\phi^q(x)=\phi^q(1-x)$, and for the gluon function 
$\phi^g(x)=-\phi^g(1-x)$.

  The diagrams of the $\eta'-g^*-g$ transition amplitude are shown
in Fig.1 and Fig.2. Note that we do not include the diagrams with
s-channel gluon exchange between the $g^*g$ and $q\bar{q}$ (gg) here,
because they are color suppressed in the case that $q\bar{q}$ (gg) will
hadronize into mesons. According to the symmetry propertis of the gluon
wave function, it is easily found that Fig.2(c) does not contribute 
to this amplitude.

  The contribution of the quark wave function to the $\eta'-g^*-g$ vertex
(see Fig.1) is 
\begin{eqnarray}
    T^q&=&c\int_0^1 dx\frac{1}{4N} \phi^q(Q^2,x)Tr[\gamma_5
         / {\hskip -2.4mm p}T^q_H],\\
      c&=&c_qf_qsin\phi +c_sf_scos\phi,\nonumber
\end{eqnarray}
where $N$ is the color number, $T^q_H$ is the hard amplitude of
the quark parton contribution. $c_q=\sqrt{2}$, $c_s=1$, $f_q$ and 
$f_s$ are the decay constant of $\eta'_q$ and $\eta'_s$ 
respectively. The gluon wave function contribution is (Fig.2)
\begin{equation}
  T^g=c\int_0^1 dx \frac{1}{4N}\phi^g(Q^2,x)\cdot i\varepsilon
      ^{\alpha\beta\rho\sigma}q_{\alpha}l_{\beta}
       (T^g_H)_{\rho\sigma}^{\mu\nu},
\end{equation}
here $q=(p+q_2)/Q$, $l=(p-q_2)/Q$, and $Q^2=q_1^2$. $Q^2$ can be 
chosen as the evolution scale of the wave functions.

  The $\eta'g^*g$ transition form fator $F_{\eta'g^*g}$ can be
defined through
\begin{equation}
    T^q+T^g= F_{\eta'g^*g}(q_1^2=Q^2,q_2^2=0,m_{\eta'}^2)
             \cdot \delta^{ab}\varepsilon^{\alpha\beta\mu\nu}
             q_{1\alpha}q_{2\beta}
\end{equation}
The indices $a$ and $b$ are $SU(3)_c$ generator indices. Calculating
the Feynman graphs shown in Fig.1 and 2, we can obtain
\begin{eqnarray}
  &&~~F_{\eta'g^*g}(q_1^2=Q^2,q_2^2=0,m_{\eta'}^2)\nonumber\\
  &&=~4\pi\alpha_s(Q^2)(c_qf_qsin\phi +c_sf_scos\phi)\nonumber\\
  &&\left\{\frac{1}{2N}\int_0^1dx\phi^q(Q^2,x)\left[\frac{1}
  {(x^2-x)p^2+(1-x)q_1^2}+(x\leftrightarrow (1-x))\right]\right.
\nonumber\\
  &&\left.-\frac{1}{2Q^2}\int_0^1dx\phi^g(Q^2,x)\left[
  \frac{(1+x(1-x))p^2-xq_1^2}
  {(x^2-x)p^2+(1-x)q_1^2}+(x\leftrightarrow (1-x))\right]\right\}.
\end{eqnarray}
The decay constants $f_q$, $f_s$ and the mixing angle $\phi$ have 
been constrained from the avialable experimental data,
$f_q=(1.07\pm 0.02)f_{\pi}$, $f_s=(1.34\pm 0.06)f_{\pi}$,
$\phi=39.3^0\pm 1.0^0$ \cite{16}. The free parameters exist
in the evolution functions $\phi^q(Q^2,x)$ and $\phi^g(Q^2,x)$ (see eq.(2)).
They are $B^q_n$ and $B^g_n$, $n=2,4,\cdots$. The fit to the 
experimental data of the $\eta'-\gamma$ transition form factor shows
that $\phi^q(Q^2,x)$ should not be much different from the asymptotic
form $\phi_{AS}(x)=6x(1-x)$, i.e., the parameters $B^q_n$ and 
$\rho^g_n B^g_n$ should be small enough ($\rho^g_n$ are not free
parameters, see Appendix A). The parameters $\gamma_+^n$ and 
$\gamma_-^n$ are negative and their absolute values increase with $n$. 
Consequently it is a good
approximation to consider only the first one or two terms in the 
expansion of the wave function $\phi^q(Q^2,x)$ and
$\phi^g(Q^2,x)$. In this work we only take into account
the first term in the expansion of the wave functions. We keep
$\mid B^q_2\mid$ and $\mid \rho^g_2 B^g_2\mid <0.1$ in order to keep
the constriant of the experiment data of the $\eta'-\gamma$ transition
form factor. Because $\rho^g_2=-0.05$, we take $\mid B^g_2\mid <2.0$.

If we adopt the constraints $\mid B^q_2\mid<0.1$, $\mid B^g_2\mid <2.0$, 
we find that the
contribution of the gluon wave function is very small, it can be neglected.
Because the gluonic contribution is so small, the dominant contribution
comes from the asymptotic quark wave function. The dependence of the QCD 
transition form factor $F_{\eta' g^* g} $ on the free parameters $B^q_2$
and $B^g_2$ is extremely weak. As an example, we present the functional form
of $F_{\eta'g^*g}$ 
with $B^q_2=0$, $B^g_2=2.0$ in Fig.3. 
We can see from the figure, after taking into account the gluonic wave 
function, the total result is not different greatly from the quark wave
function contribution.

In fig.4 we compare our result (solid curve in Fig.4) with what were used 
in the literatures: i) in \cite{5,9,10} the $\eta'$-gluon transition form
factor
is taken as $\frac{H(0,0,m_{\eta'}^2)}{(q_1^2/m_{\eta'}^2-1)}$, where
$H(0,0, m_{\eta'}^2)$ is a phenomenonolgical parameter which should be
extracted from the branching ratio of $\psi\to\eta'\gamma$, $H(0,0, m_{\eta'}^2)\approx 1.8 GeV^{-1}$ (see the dashed curve in Fig.4); ii) in \cite{4,8} the
form factor is taken as $\sqrt{3} \alpha_s(Q^2)/(\pi f_{\pi})$
(see the dot-dashed curve in Fig.4).
Our result is very close to $\frac{1.8GeV^{-1}}{(q^2/m_{\eta'}^2-1)}$. If we
take $H(0,0, m_{\eta'}^2)=1.7 GeV^{-1}$, the curve is even closer to our
result. In \cite{3,5,9,10} the branching fraction of $\psi\to\eta'\gamma$
is needed to extracted the parameter $H(0,0,m_{\eta'}^2)$; here we do not 
need to extract the parameter $H(0,0, m_{\eta'}^2)$. The structure of $\eta'$
can determine the behavior of the form factor $F_{\eta'g^*g}$ completely.
It is also possible to calculate $F_{\eta'g^*g^*}$ by using the same method.
Certainly a large number of experiments such as $pp\to \eta'x$, $pp\eta'$,
etc., are needed to test the behavior of the $\eta'$-gluon transition 
form factor. The behavior of the $\eta'$-gluon transition form factor is not
only important in investigating the dynamics of $\eta'$ production from
B decays, but also in determining the structure of $\eta'$. Thus such kinds
of experiments are urgently needed.

The summary: we calculated the QCD transition form factor 
$F_{\eta'g^*g}(Q^2,0,m_{\eta'}^2)$ by using the wave function of $\eta'$
which is abtained by solving the evolution equation \cite{17}. We included the 
gluonic wave function in our calculation. We find within the possible free
parameter region, the gluonic contribution is small, and the QCD transition
form factor $F_{\eta'g^*g}$ does not depend on the free parameters greatly.

\vspace{1cm}
This work is supported by Monbusho Fund 10098178-00.
One of us (MZY) thanks Japan Society for the 
Promotion of Science (JSPS) for financial support.

\newpage

\begin{center} Appendix A \end{center}

The parameters $\gamma ^n$ and $\rho_n$ in eq.(2):
$$\gamma^n_{\pm}=\frac{1}{2}\{\gamma^n_{qq}+\gamma^n_{gg}\pm
            \sqrt{(\gamma^n_{qq}-\gamma^n_{gg})^2+4\gamma^n_{gq}\gamma^n_{qg}}
            \}, \eqno(A1)$$
$$\gamma^n_{qq}=\frac{C_F}{\beta}\left\{\frac{2}{(n+1)(n+2)}-1-
               4\sum\limits_{j=2}^{n+1}\frac{1}{j}\right\}, \eqno(A2)$$
where $C_F=\frac{4}{3}$, $\beta=\frac{11N-2n_f}{3}$, $N$ is the number of color,$n_f$ is the number of the active quarks.
$$\gamma^n_{gq}=\frac{n_f}{\beta}\frac{2}{(n+1)(n+2)}, \eqno(A3)$$
$$\gamma^n_{qg}=\frac{C_F}{\beta}\frac{n(n+3)}{(n+1)(n+2)}, \eqno(A4)$$
$$\gamma^n_{gg}=\frac{4N}{\beta}\left\{\frac{2}{(n+1)(n+2)}-
               \sum\limits_{j=2}^{n+1}\frac{1}{j}-\frac{1}{12}-
               \frac{n_f}{6N} \right\}, \eqno(A5)$$
where $n\geq 1$.
$$P_n=\frac{\gamma^n_{+}-\gamma^n_{qq}}{\gamma^n_{+}-\gamma^n_{-}}, ~~~~~
  Q_n=\frac{\gamma^n_{qg}}{\gamma^n_{+}-\gamma^n_{-}},  \eqno(A5)$$
$$\rho^g_n=-\frac{1}{6}\frac{Q_n}{1-P_n}, ~~~~~\rho^q_n=6\frac{P_n}{Q_n}.
   \eqno(A6)$$

\newpage

\begin{center}  Figure Captions \end{center}

  Fig1. The quark content contribution to the $\eta'g^*g$ transition 

\vspace{1cm}

  Fig2. The gluonic content contribution to the $\eta'g^*g$ transition

\vspace{1cm}

  Fig3. The QCD transition form factor 
        $F_{\eta'g^*g}(Q^2,0,m^2_{\eta'})$. The dot-dashed curve is 
        the contribution of the quark wave function, the dashed
        curve is the gluonic contribution ($B^2_2=0$, $B^g_2=2.0$),
        and the solid one is the total contribution.

\vspace{1cm}

  Fig4. The comparison of $F_{\eta' g^* g}$ which we obtain with others
        used in literatures: i) The solid curve is the result calculated
        based on the $\eta'$ wave function; ii) The dashed curve is for
        $\frac{H(0,0,m_{\eta'}^2)}{(q_1^2/m_{\eta'}^2-1)}$ with 
        $H(0,0,m_{\eta'}^2)=1.8GeV^{-1}$; iii) The dot-dashed one is
        $\sqrt{3}\alpha_s(Q^2)/(\pi f_{\pi})$

\newpage

\begin{figure}
\setlength{\unitlength}{0.13in} 
\begin{center}
\begin{picture}(35,10)
\put(11,6){\line(-1,0){4}}
\put(11,3){\line(-1,0){4}}
\put(7,6){\line(0,-1){3}}

\multiput(6.5,5.8)(-0.5,0.5){8}{\oval(1,1)[t]}
\multiput(6.0,6.3)(-0.5,0.5){8}{\oval(1,1)[r]}
\multiput(6.5,3.2)(-0.5,-0.5){8}{\oval(1,1)[b]}
\multiput(6.0,2.7)(-0.5,-0.5){8}{\oval(1,1)[r]}

\put(2,7){$q_1$}
\put(3.5,7.5){\vector(1,-1){1.1}}
\put(2,1){$q_2$}
\put(3.5,1.5){\vector(1,1){1.1}}
\put(8,7.5){$xp$}
\put(8.5,7){\vector(1,0){1}}
\put(7,1){$(1-x)p$}
\put(8.5,2){\vector(1,0){1}}
\put(1,10){$g^*$}
\put(1.3,-1.5){$g$}
\put(12,4){$\eta'$}
\put(27,6){\line(-1,0){4}}
\put(27,3){\line(-1,0){4}}
\put(23,6){\line(0,-1){3}}

\multiput(22.5,6.0)(-0.5,-0.5){12}{\oval(1,1)[t]}
\multiput(22.5,6.0)(-0.5,-0.5){11}{\oval(1,1)[l]}
\multiput(22.5,3.0)(-0.5,0.5){12}{\oval(1,1)[b]}
\multiput(22.5,3.0)(-0.5,0.5){11}{\oval(1,1)[l]}
\end{picture}
\end{center}
\caption{}
\end{figure}

\vspace{1cm}

\begin{figure}
\setlength{\unitlength}{0.13in} 
\begin{center}
\begin{picture}(35,10)
\multiput(4.5,6)(0.6,0){7}{\oval(1,1)[t]}
\multiput(4.8,6)(0.6,0){6}{\oval(0.4,0.4)[b]}
\multiput(4.5,3.2)(0.6,0){7}{\oval(1,1)[b]}
\multiput(4.8,3.2)(0.6,0){6}{\oval(0.4,0.4)[t]}
\multiput(4,5.5)(0,-0.6){4}{\oval(1,1)[r]}
\multiput(4,5.2)(0,-0.6){3}{\oval(0.4,0.4)[l]}
\put(4,6){\circle*{0.5}}
\put(4,3.2){\circle*{0.5}}

\multiput(3.5,6.0)(-0.5,0.5){8}{\oval(1,1)[t]}
\multiput(3.0,6.5)(-0.5,0.5){8}{\oval(1,1)[r]}
\multiput(3.5,3.2)(-0.5,-0.5){8}{\oval(1,1)[b]}
\multiput(3.0,2.7)(-0.5,-0.5){8}{\oval(1,1)[r]}

\put(4,-2){$(a)$}
\put(-1,7){$q_1$}
\put(1,7.5){\vector(1,-1){1.1}}
\put(-1,1){$q_2$}
\put(1,1.5){\vector(1,1){1.1}}
\put(5,7.5){$xp$}
\put(5.5,7){\vector(1,0){1}}
\put(4,1){$(1-x)p$}
\put(5.5,2){\vector(1,0){1}}
\put(1,10){$g^*$}
\put(0.5,-1.5){$g$}
\put(9,4){$\eta'$}

\multiput(20.5,6)(0.6,0){7}{\oval(1,1)[t]}
\multiput(20.8,6)(0.6,0){6}{\oval(0.4,0.4)[b]}
\multiput(20.5,3.2)(0.6,0){7}{\oval(1,1)[b]}
\multiput(20.8,3.2)(0.6,0){6}{\oval(0.4,0.4)[t]}
\multiput(20,5.5)(0,-0.6){4}{\oval(1,1)[r]}
\multiput(20,5.2)(0,-0.6){3}{\oval(0.4,0.4)[l]}
\put(20,6){\circle*{0.5}}
\put(20,3.2){\circle*{0.5}}

\multiput(19.5,6.0)(-0.5,-0.5){12}{\oval(1,1)[t]}
\multiput(19.5,6.0)(-0.5,-0.5){11}{\oval(1,1)[l]}
\multiput(19.5,3.2)(-0.5,0.5){12}{\oval(1,1)[b]}
\multiput(19.5,3.2)(-0.5,0.5){11}{\oval(1,1)[l]}

\put(20,-2){$(b)$}

\multiput(36.1,6)(0.6,0){7}{\oval(1,1)[t]}
\multiput(35.8,6)(0.6,0){7}{\oval(0.4,0.4)[b]}
\multiput(36.1,3.2)(0.6,0){7}{\oval(1,1)[b]}
\multiput(35.8,3.2)(0.6,0){7}{\oval(0.4,0.4)[t]}

\put(34,4.6){\circle*{0.8}}

\multiput(35.5,6.0)(-0.5,-0.5){12}{\oval(1,1)[t]}
\multiput(35.5,6.0)(-0.5,-0.5){11}{\oval(1,1)[l]}
\multiput(35.5,3.2)(-0.5,0.5){12}{\oval(1,1)[b]}
\multiput(35.5,3.2)(-0.5,0.5){11}{\oval(1,1)[l]}

\put(37,-2){$(c)$}

\end{picture}
\end{center}
\caption{}
\end{figure}

\vspace{1cm}

\begin{figure}
    \begin{center}
      \epsfig{file=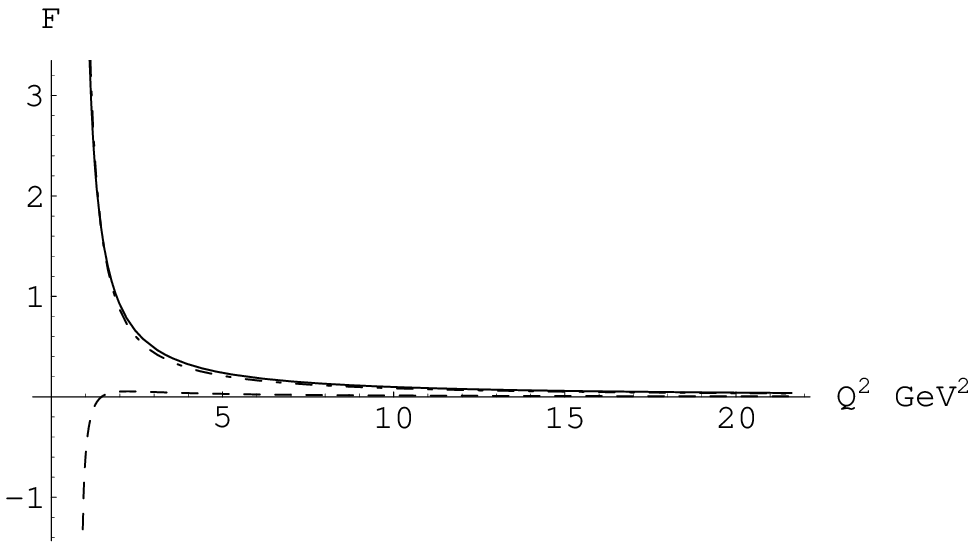, width=10cm}
      \caption{}
    \end{center}
\end{figure}

\vspace{1cm}

\begin{figure}
    \begin{center}
      \epsfig{file=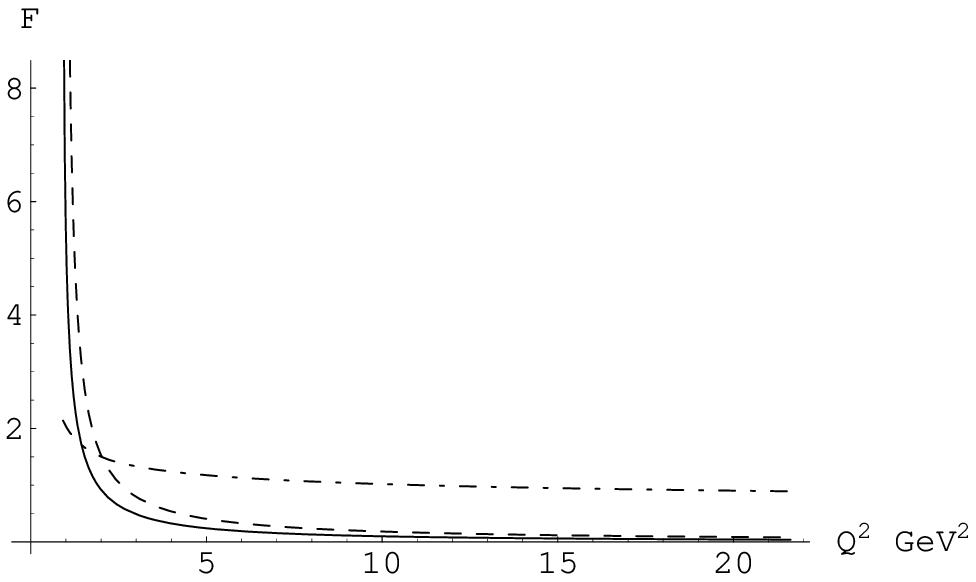, width=10cm}
      \caption{ }
    \end{center}
\end{figure}

\end{document}